\documentclass[aip,
 sd,
 amsmath,amssymb,
 reprint,
]{revtex4-1}

\usepackage{graphicx}
\usepackage{dcolumn}
\usepackage{bm}
\usepackage{mathtools}
\usepackage{xcolor}

\begin{document}

\preprint{AIP/123-QED}

\title{Competition of noise and collectivity in global cryptocurrency trading: route to a self-contained market.}

\author{Stanis\l aw Dro\.{z}d\.{z}}
\altaffiliation{Author to whom correspondence should be addressed. Electronic address: stanislaw.drozdz@ifj.edu.pl}
\affiliation{Complex Systems Theory Department, Institute of Nuclear Physics Polish Academy of Sciences, ul. Radzikowskiego 152, 31--342 Krak\'ow, Poland.}
\affiliation{Faculty of Computer Science and Telecommunication, Cracow University of Technology, ul. Warszawska 24, 31--155 Krak\'ow, Poland.}
\author{Ludovico Minati}
\affiliation{Complex Systems Theory Department, Institute of Nuclear Physics Polish Academy of Sciences, ul. Radzikowskiego 152, 31--342 Krak\'ow, Poland.}
\author{Pawe\l{} O\'swi\c{e}cimka}
\affiliation{Complex Systems Theory Department, Institute of Nuclear Physics Polish Academy of Sciences, ul. Radzikowskiego 152, 31--342 Krak\'ow, Poland.}
\author{Marek Stanuszek}
\affiliation{Faculty of Computer Science and Telecommunication, Cracow University of Technology, ul. Warszawska 24, 31--155 Krak\'ow, Poland.}
\author{Marcin W\c{a}torek }
\affiliation{Complex Systems Theory Department, Institute of Nuclear Physics Polish Academy of Sciences, ul. Radzikowskiego 152, 31--342 Krak\'ow, Poland.}
\affiliation{Faculty of Computer Science and Telecommunication, Cracow University of Technology, ul. Warszawska 24, 31--155 Krak\'ow, Poland.}
         
\date{\today}

\begin{abstract}
Cross-correlations in fluctuations of the daily exchange rates within the basket of the 100 highest-capitalization cryptocurrencies over the period October 1, 2015, through March 31, 2019, are studied. The corresponding dynamics predominantly involve one leading eigenvalue of the correlation matrix, while the others largely coincide with those of Wishart random matrices. However, the magnitude of the principal eigenvalue, and thus the degree of collectivity, strongly depends on which cryptocurrency is used as a base. It is largest when the base is the most peripheral cryptocurrency; when more significant ones are taken into consideration, its magnitude systematically decreases, nevertheless preserving a sizable gap with respect to the random bulk, which in turn indicates that the organization of correlations becomes more heterogeneous. This finding provides a criterion for recognizing which currencies or cryptocurrencies play a dominant role in the global crypto-market. The present study shows that over the period under consideration, the Bitcoin (BTC) predominates, hallmarking exchange rate dynamics at least as influential as the US dollar. Even more, the BTC started dominating around the year 2017, while further cryptocurrencies, like the Ethereum (ETH) and even Ripple (XRP), assumed similar trends. At the same time, the USD, an original value determinant for the cryptocurrency market, became increasingly disconnected, and its related characteristics eventually started approaching those of a fictitious currency. These results are strong indicators of incipient independence of the global cryptocurrency market, delineating a self-contained trade resembling the Forex.
\end{abstract}

                            
\maketitle

\begin{quotation}
The coexistence of noise and collectivity, with elements of competition, arguably constitutes the most salient feature of complexity. The related characteristics are well-evident in the financial markets wherein, at the level of individual time-series, they are quantified in terms of heavy-tailed distributions of returns and clustering of activity, resulting in non-linear long-range temporal correlations, and giving rise to multiscaling effects. When the relationships between multiple assets are considered via cross-correlations, an even richer spectrum of behaviors is revealed. Owing to the direct relevance in optimal portfolio construction, matrix formalism attains particular relevance. It is well-established that in the mature global markets, the cross-correlations are dominated by noise effects and the related characteristics can thus be recapitulated via predictions from appropriate ensembles of random matrices. At the same time, normally at least one outstandingly large eigenvalue, vastly exceeding the random ensemble boundaries, is found; together with the corresponding eigenvector, this reflects the synchronous, or collective, drift of the entire market. The emerging cryptocurrency trading provides a unique opportunity for exploring the development of a currency market in its early stages, which is the focus of the present study.
\end{quotation}

\section{Introduction}
Only a relatively short time has elapsed since the introduction of the first fiat-to-bitcoin exchange (Mt. Gox), in July 2010, and of the first rules-free decentralized marketplace (Silk Road), in February 2011~\cite{berentsen2018}. Nevertheless, evidence has already accumulated indicating that, according to the crucial complexity characteristics in the return distribution, temporal correlations, and multiscaling effects~\cite{kwapien2012}, the Bitcoin market has become factually indistinguishable from the conventional mature markets, including, in particular, the foreign exchange (Forex) market \cite{drozdz2018}. Furthermore, it has been postulated that the other cryptocurrencies would follow a similar trajectory, eventually heralding the emergence of an entirely new market, analogous to the global Forex market, wherein the same are traded in a largely self-contained manner. Preliminary evidence confirming this trend has already been provided \cite{drozdz2019}.\par

The spectacular growth of the global cryptocurrency market during the second half of the year 2017, propelled by the so-called Initial Coin Offer (ICO) bubble \cite{aste2019}, during which the number of traded cryptocurrencies doubled from 700 to 1400 and the total market capitalization rose to almost 800 billion USD, brought mainstream media~\cite{WSJ} and research attention to the topic\cite{corbet2019}. This mania period eventually lead to a crash in Jan 2018 \cite{gerlach2018}, but, since then, cryptocurrencies have stabilized their role as a component of the modern financial markets, especially after the introduction of the first futures contract on the Bitcoin price. Consequently, research uncovering the internal correlations within the cryptocurrency market~\cite{bariviera2018,stosic2018,bouri2019,zieba2019,polonikov2020} as well as its relationships with the mature markets~\cite{Szetela2016,corbet2018,corelli2018,ji2018,kristj2019}, began to accumulate. In particular, the possible use of Bitcoin as a `hedge or safe haven' for currencies~\cite{urquhart2019}, gold and commodities~\cite{shahzad2019}, or stock markets~\cite{shahzad2019a,wang2019}, has been considered.\par

Motivated by such rapid developments, here a more systematic analysis is conducted via quantifying cross-correlations among a comprehensive set of highest-capitalization cryptocurrencies, establishing parallels with the characteristics of the stock market~\cite{laloux1999,plerou1999,drozdz2001}, particularly the Forex~\cite{drozdz2007,gorski2008}, as regards the coexistence of noise and collectivity. 

\section{Data specification}
\begin{figure}
\includegraphics[scale=0.32]{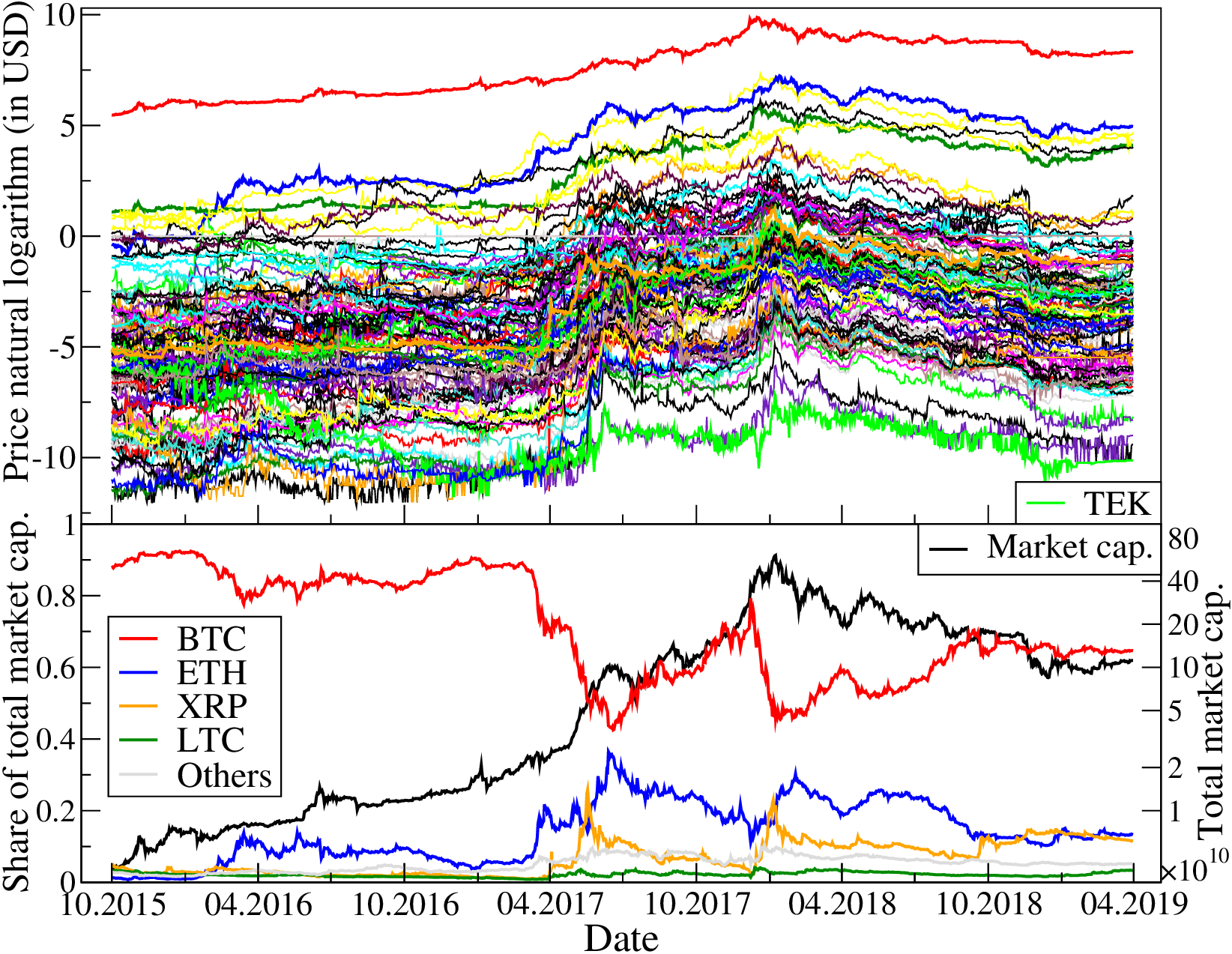} 
\caption{Time-development of the prices of 100 highest-capitalization cryptocurrencies expressed in USD and based on daily recordings from October 1, 2015, through March 31, 2019 (top), and time-developments of the corresponding total capitalization (black line) and the fractions of Bitcoin (BTC), Ethereum (ETH), Ripple (XRP), Litcoin (LTC) and TEKcoin (TEK) (bottom).}
\label{fig:crypto-all}
\end{figure}

The dataset consists of $T=1,278$ daily records pertaining to the $n=100$ highest-capitalization cryptocurrencies (all of them are explicitly listed in Section III.D), covering the period from October 1, 2015, through March 31, 2019; it was drawn from \verb+CoinMarketCap+~\cite{coinmarket}, a recognized website tracking the price and capitalization of cryptocurrencies. This dataset thus covers the major up-down crypto-market trend reversal, which occurred between the years 2017-2018. All prices, expressed in USD, were determined by averaging over selected major crypto-market exchanges (273 in 2019), including crypto-crypto exchanges, and weighted by volume. The market capitalization was calculated by multiplying the price with the total supply. Fluctuations in the logarithm of the prices of all cryptocurrencies under consideration are shown in Fig.~\ref{fig:crypto-all} (top), alongside fluctuations in the total crypto-market capitalization level (bottom). The capitalization can be seen to exceed more than half a trillion USD by the end of the year 2017. Paralleling the overall market decline, it then dropped but started stabilizing at a level still noticeably beyond 100 billion USD, thus exceeding the capitalization of several conventionally-recognized stock and commodity markets. It appears noteworthy that, until early 2017, 90$\%$ of this capitalization could be ascribed to the Bitcoin (BTC). Subsequently, other cryptocurrencies began playing an increasing role, spearheaded by the Ethereum (ETH), which by the mid-2017 reached a capitalization level approaching that of the BTC. Another notably-capitalized emerging cryptocurrency is the Ripple (XRP), which, in exceeding 50 billion USD, reached about the same level as the ETH by the end of the study period. The fourth cryptocurrency by rank is the Litcoin (LTC), whose capitalization stabilized around 10 billion USD by the same time. The remaining 96 cryptocurrencies in the basket (with TEKcoin having the lowest capitalization) in total account for not more than 10$\%$ of the entire market capitalization.\par

\section{Correlation matrix} 

The interdependencies among the fluctuations in the exchange rates were addressed by means of the correlation matrix, in calculating which any of the cryptocurrencies can be taken as a base~\cite{kwapien2009,basnarkov2019}. From the resulting price series $P^{(\alpha)}_i(t)$ of length $T$ a correlation matrix was, therefore, separately determined for each of the $n$ base cryptocurrencies $\alpha$ (thus corresponding to denominators), yielding $n$ matrices of size $(n-1) \times (n-1)$, ${\bf C^{(\alpha)}} \equiv [C^{(\alpha)}_{ij}]$, where $i, j = 1, ..., n-1$ and $\alpha$ denotes the base cryptocurrency $(\alpha = 1, ..., n)$. In this context, for the series of log returns $G^{(\alpha)}_i (t; \tau) = \log (P^{(\alpha)}_i (t + \tau)) - \log (P^{(\alpha)}_i(t))$, where $\tau$ is a time-lag, which in the present case equals 1 day. The correlation matrix is defined via the auxiliary matrix $\bf M^{(\alpha)}$ of normalized log returns, 
\begin{equation}
g^{(\alpha)}_i (t; \tau) = {{G^{(\alpha)}_i (t; \tau)-\langle G^{(\alpha)}_i (t; \tau) \rangle_T} \over {\sigma(G^{(\alpha)}_i)}},
\label{g}
\end{equation}
where $\sigma(G)$ denotes the standard deviation of $G$. Such a normalization of each series independently is commonly applied for taking care of the heteroskedasticity effects. Taking $(n-1)$ time-series $g^{(\alpha)}_i(t; \tau)$, $g^{(\alpha)}_i(t+\tau; \tau),..., g^{(\alpha)}_i(t+(T-1)\tau; \tau$) of length $T$, one can construct an $(n-1) \times T$ rectangular matrix ${\bf M^{(\alpha)}}$. Finally, the symmetric correlation matrix is arrived at with 
\begin{equation}
{\bold C}^{(\alpha)} = {1 \over T} {\bold M}^{(\alpha)} {\tilde {\bold M}}^{(\alpha)}
\label{C}
\end{equation}
where superscript $\sim$  denotes matrix transposition. By construction, each correlation matrix possesses $n-1$ real eigenvalues $(\lambda^{(\alpha)}_i)$ and corresponding eigenvectors $(v^{(\alpha)}_{ij} \equiv {\bf v}^{(\alpha)}_i)$:
\begin{equation}
{\bf C}^{(\alpha)} {\bf v}_i^{(\alpha)} = \lambda^{(\alpha)} {\bf v}_i^{(\alpha)}.
\label{eigenequation}
\end{equation}
The eigenvalues are enumerated in monotonically increasing order, $0 \le \lambda^{(\alpha)}_{1} \le \lambda^{(\alpha)}_{2} \le$...$\le \lambda^{(\alpha)}_{(n-1)}$. The trace $\textrm{Tr} {\bf C}^{(\alpha)} = n-1$, i.e., equal to the number of time-series included in ${\bf M}^{(\alpha)}$. Via their expansion coefficients $v^{(\alpha)}_{ij}$, the eigenvectors ${\bf v}^{(\alpha)}_i$ can be associated to the corresponding eigensignals $z^{(\alpha)}_i(t)$ generated from the original time-series as
\begin{equation}
z^{(\alpha)}_i(t)= \sum_{j=1}^{n-1} v^{(\alpha)}_{ij} g^{(\alpha)}_j(t),
\label{eigensignal}
\end{equation}
thus prescribing the corresponding portfolio's returns~\cite{markowitz}.

\subsection{Wishart limit case}

Each matrix ${\bold C}^{(\alpha)}$ was considered individually, the focus being on the identification of potentially non-random instances. Assuming that a purely random case, wherein no correlations are present, corresponds to the Wishart~\cite{wishart1928} ensemble of random matrices $\bf W$, the following eigenvalue density is arrived at:
\begin{equation}
\rho_W(\lambda)={1 \over N} \sum_{k=1}^N \delta(\lambda - \lambda_k) = {Q \over 2 \pi \sigma_{\textrm{W}}^2} {\sqrt{(\lambda_{+}-\lambda)(\lambda-\lambda_{-})} \over \lambda},
\label{rhoW}
\end{equation}

\begin{equation}
\lambda_{\pm} = \sigma_{\textrm{W}}^2 (1 + 1/Q \pm 2 \sqrt{1 \over Q}),
\label{lambdaW}
\end{equation}
where $\delta(x)$ denotes the Dirac delta function, $\lambda\in[\lambda_-,\lambda_+]$ and $Q=T/N$. Here, $T$ and $N$ denote, respectively the length and number of time-series. These expressions, known under the names of Marchenko-Pastur distribution, hold exactly in the limit $T, N \to \infty$~\cite{marcenko1967}.

\subsection{Matrix elements}
\begin{figure}
\includegraphics[scale=0.4]{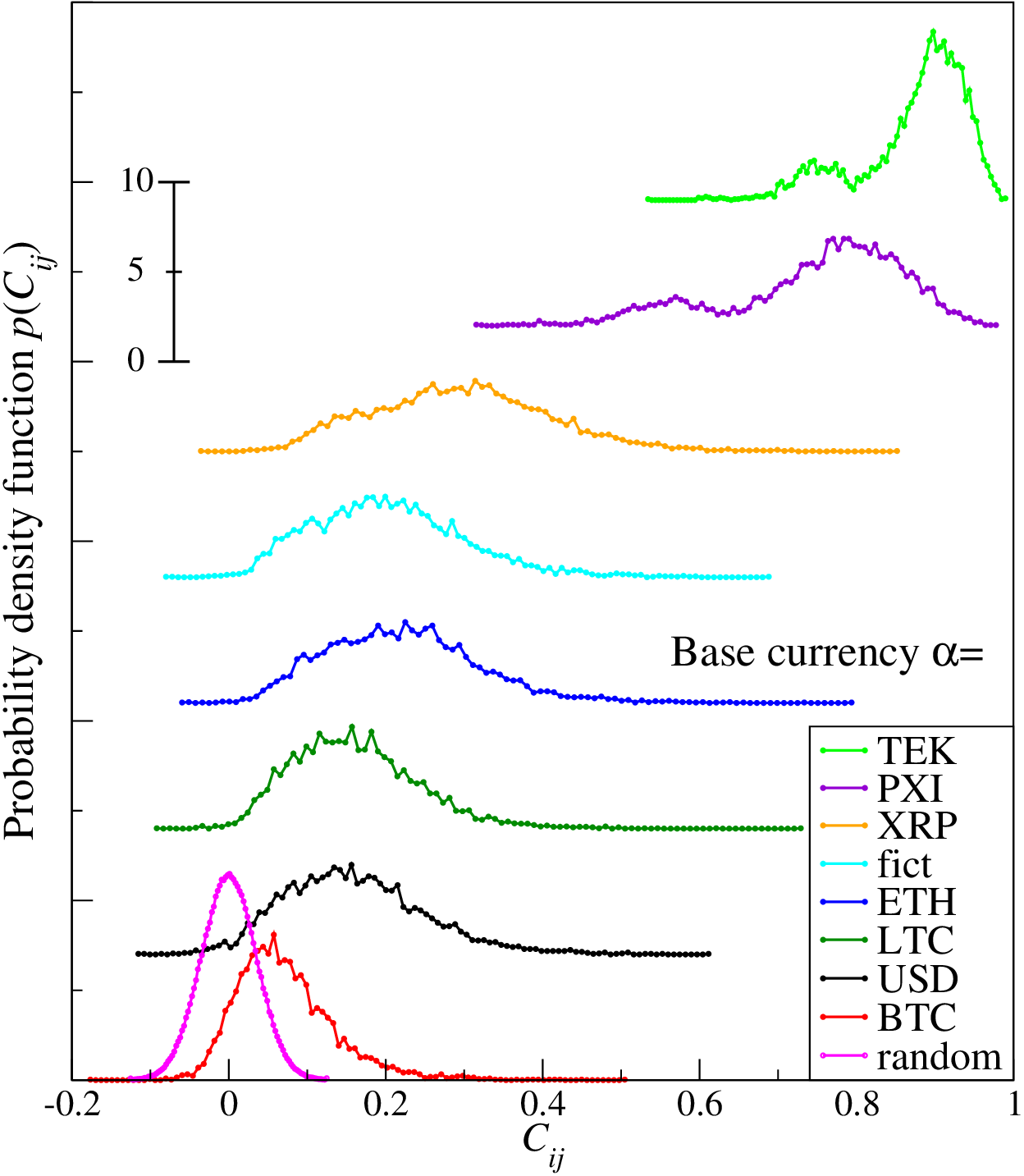} 
\caption{Distribution of the off-diagonal matrix elements of the correlation matrix $C^{(\alpha)}_{ij}$, for six selected base cryptocurrencies and for the USD. Two surrogate cases, namely an entirely random series (random) and a fictitious base currency (fict), are also presented. For visualization purposes, adjacent plots are shifted upwards.}
\label{fig:elements}
\end{figure}

Compared to an idealized random scenario wherein the correlation matrix entries are distributed according to a Gaussian curve centered around zero, correlations based on empirical data often delineate a uniform displacement of the distribution~\cite{drozdz2000}, or the appearance of fatter tails~\cite{drozdz2001}. Both these mechanisms lead to an effective reduction of the rank of the leading component of the matrix under consideration; as a consequence, a significantly larger eigenvalue $\lambda_\textrm{max}$ is expected~\cite{drozdz2002}. In the present case of cryptocurrency correlation matrices, the distribution of off-diagonal entries markedly depends on which one is chosen as the base. Representative examples, including extreme ones, are shown in Fig.~\ref{fig:elements}. They all deviate from a zero-mean Gaussian distribution, and it may seem surprising that it is the case wherein the BTC used as the reference cryptocurrency that differs the least from it. Even the original correlation matrix, wherein the crypto-exchange rates are expressed with respect to the USD, is shifted towards more positive values. The shapes of these distributions indicate that fluctuations in the exchange rates develop both positive and negative correlations, with a visible asymmetry towards positive correlation. As one moves up in Fig.~\ref{fig:elements}, i.e., the other cryptocurrencies listed are taken as a reference, the dynamics become even more positively correlated. The peak of the distribution, as well as its mean, systematically shift rightwards. In the present basket of cryptocurrencies, an extreme situation is found when all of them are expressed in terms of the TEK (TEKcoin). In this case, almost all exchange rate pairs have correlation coefficients $>0.5$, and the peak is found close to 0.9.\par

To aid visualization and understanding, two `null hypotheses' are considered, the corresponding distributions also being shown in Fig.~\ref{fig:elements}. In the first case, a pseudo-currency termed fictitious (fict), is generated so that the USD/fict exchange rate time series is represented by a sequence of uncorrelated random numbers drawn from the geometric Brownian motion GBM($\mu=0$, $\sigma=1$) process. The exchange rates of the other cryptocurrencies are then expressed in terms of such a fictitious currency, which by construction is entirely disconnected from the real-world cryptocurrency market, by using the relation: cryptocurrency/fict = cryptocurrency/USD$\times$USD/fict. However, the resulting distribution of the correlation coefficients does not differ substantially from the majority of cases wherein real cryptocurrencies are taken as a reference. In the second case, all time-series are replaced by uncorrelated random numbers retaining the same length, while resulting in zero-centered Gaussian distributions of the matrix elements.

\subsection{Eigenvalue distribution} 
\label{Eigenvalue distribution}

\begin{figure}
\includegraphics[scale=0.32]{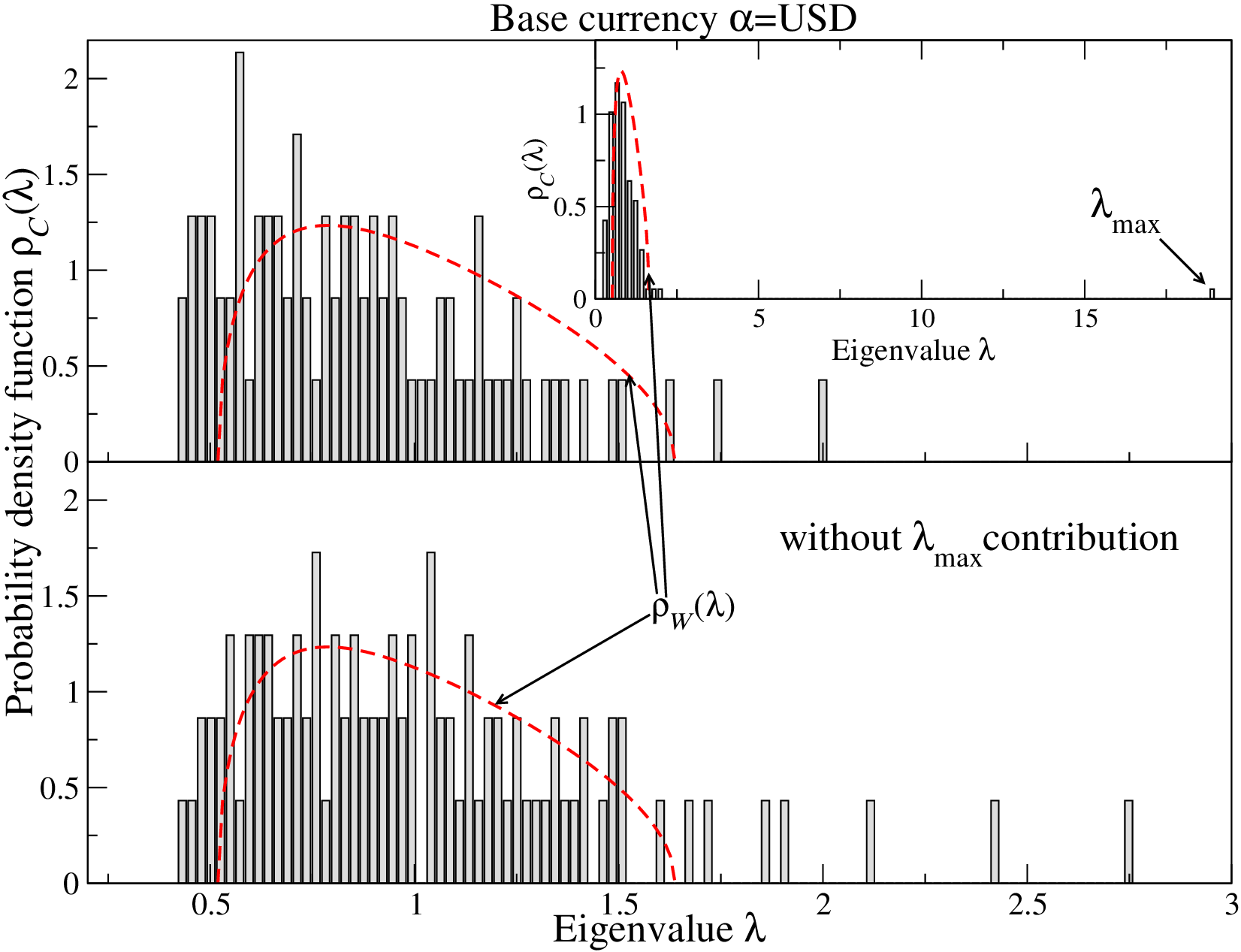}\\
\includegraphics[scale=0.32]{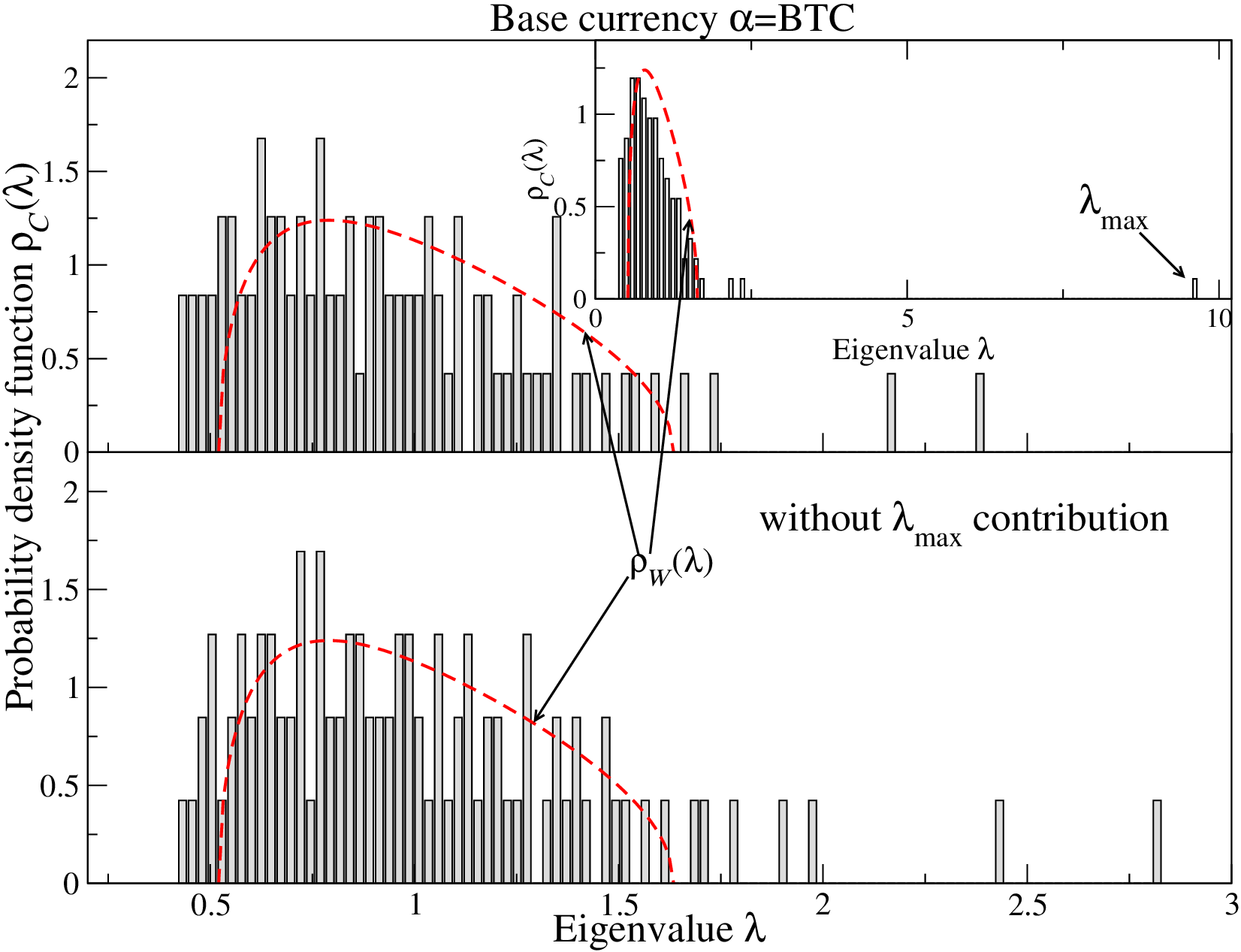}\\
\includegraphics[scale=0.32]{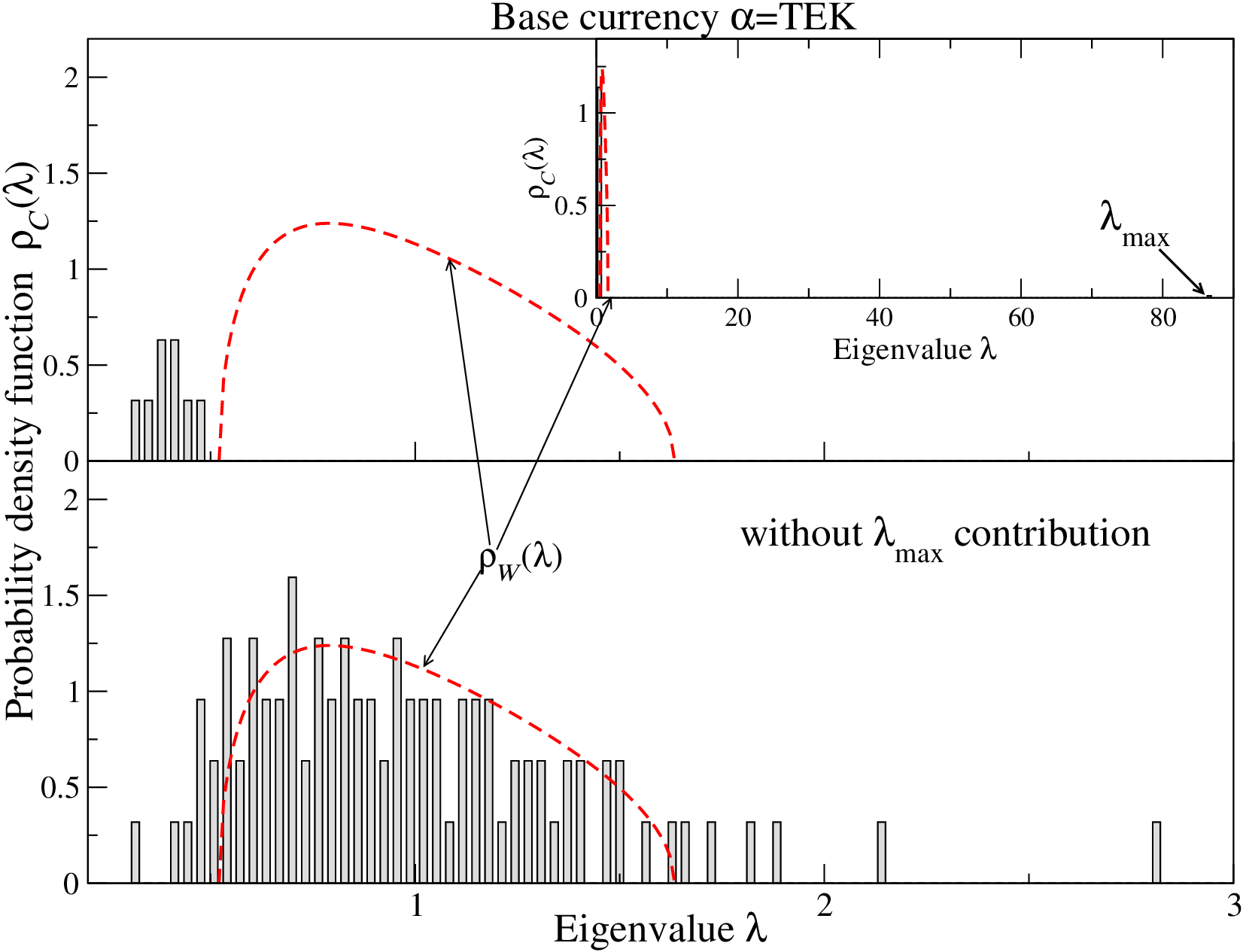}
\caption{Distribution of eigenvalues of the correlation matrix $C^{(\alpha)}_{ij}$, for the USD, BTC and TEK selected as a base for the remaining ones (histogram), together with the fitted Marchenko-Pastur distribution (dashed line). Main upper parts show the bulk of the distribution and the corresponding insets show the entire distribution. The corresponding lower parts illustrate the distributions obtained after removing the contribution of the largest eigenvalue as described in Sec.~\ref{Eigenvalue distribution}.}
\label{fig:spectra}
\end{figure}

The most informative characteristic of a correlation matrix is its eigenspectrum. Complete density of eigenspectra $\rho_C(\lambda)$ for three cases wherein the USD and the two cryptocurrency extremes, BTC and TEK, are taken as a base (these cases belong to those explicitly displayed in Fig.~\ref{fig:elements}), are shown in Fig.~\ref{fig:spectra}. From the viewpoint of the market dynamics, the most relevant quantities encompass: (i) the magnitude of the largest eigenvalue $\lambda_{\textrm{max}}$, (ii) the gap which it develops relative to the other eigenvalues (so called spectral gap), which reflects the entity of non-random correlations, and (iii) the relation of the bulk of eigenvalues to the Wishart random matrix limit, herein represented by Eq.~(\ref{rhoW}). The relative locations of the largest eigenvalues track the shifts of the peaks in the distributions of corresponding correlation matrix entries, visible in Fig.~\ref{fig:elements}. Thus, the smallest value $\lambda_{\textrm{max}} \approx 9.7$ is found for the BTC-based case, while the majority of remaining eigenvalues are bound within the limits prescribed by the Marchenko-Pastur in Eq.~(\ref{rhoW}). The USD-based case features a significantly larger value $\lambda_{\textrm{max}} \approx 19$, which in turn indicates that price changes of the cryptocurrencies expressed in this form occur more collectively. The largest value of $\lambda_{\textrm{max}}$ is recorded when TEK is used as a base.\par

\begin{figure}
\centering
\includegraphics[scale=0.32]{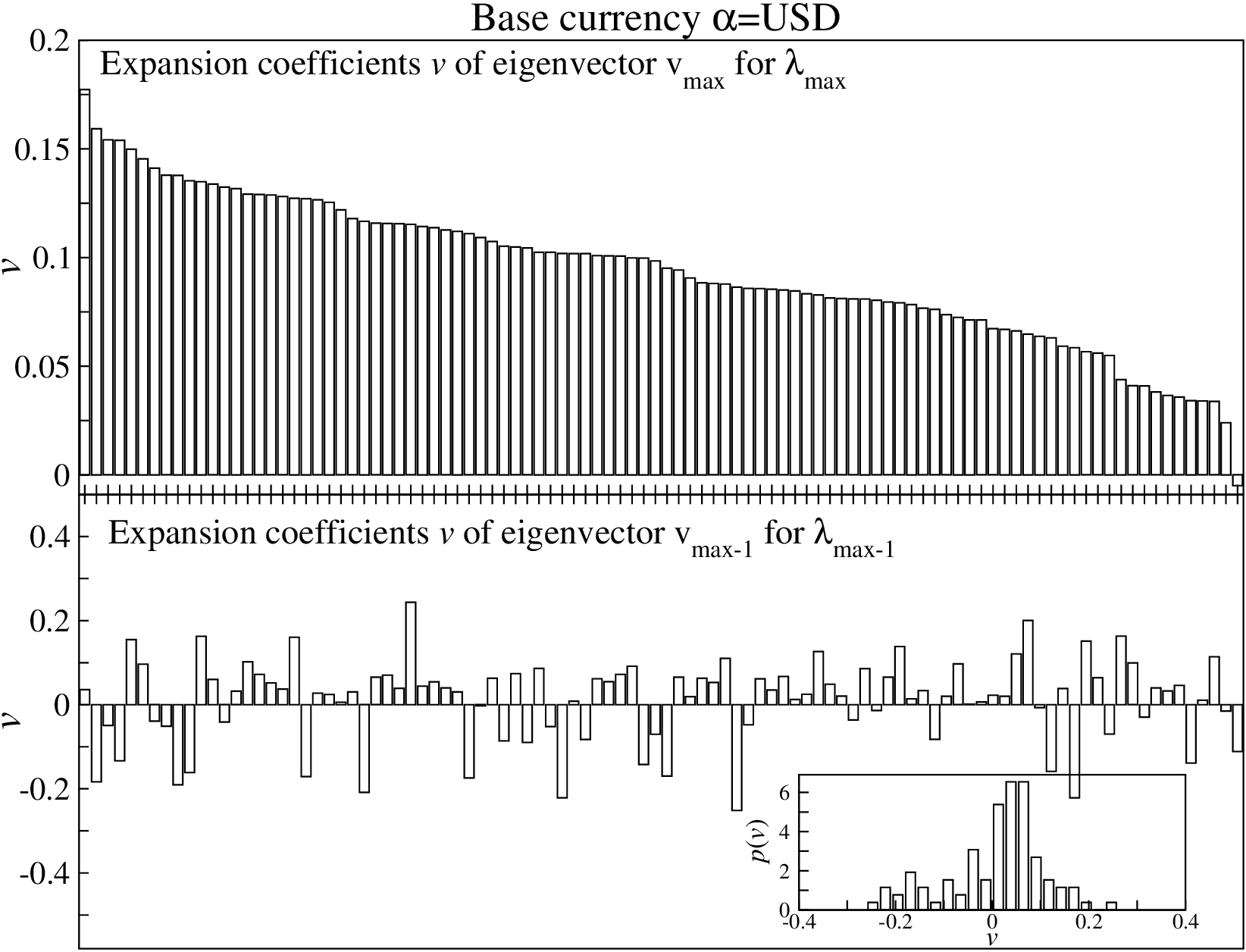}\\
\includegraphics[scale=0.32]{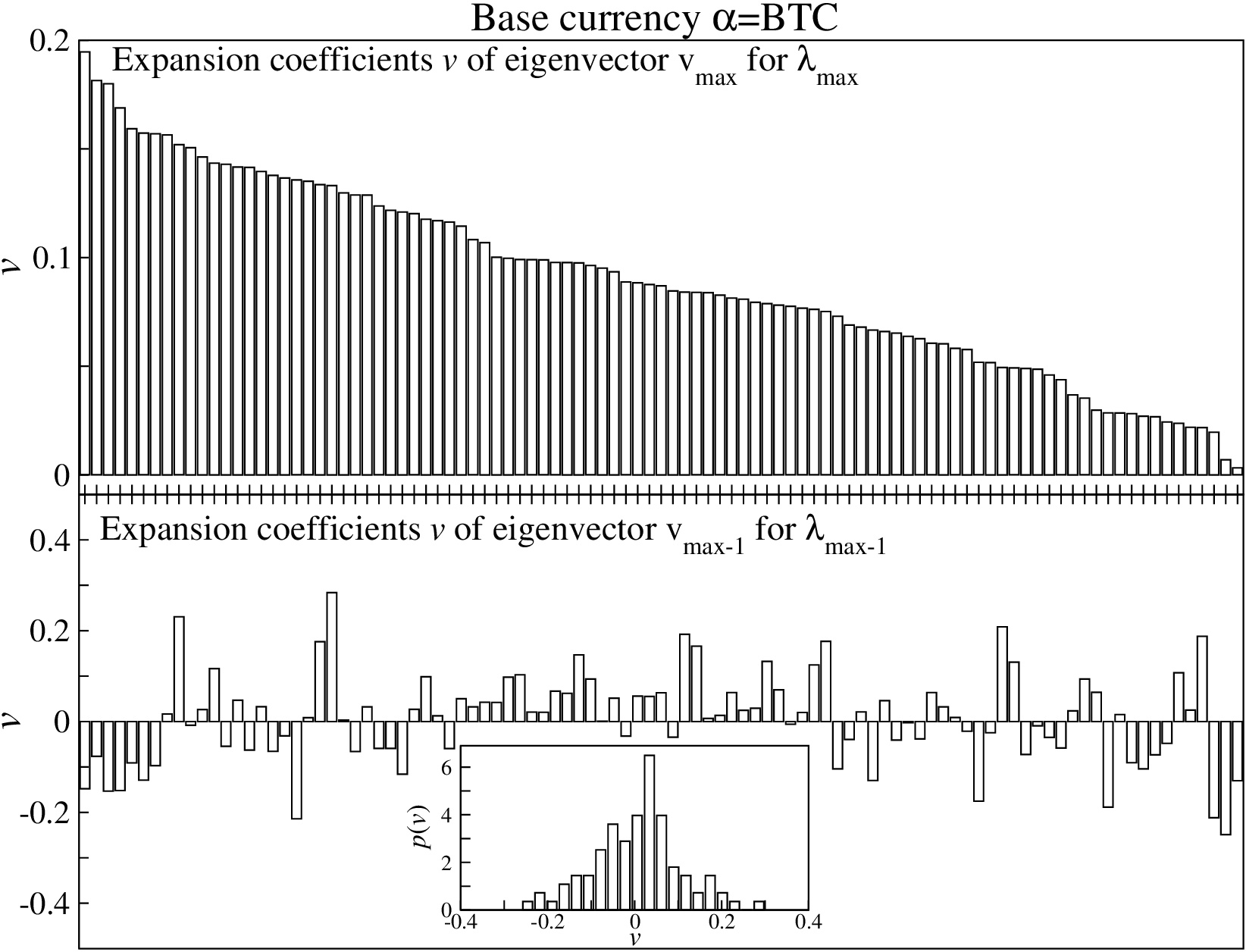}\\
\includegraphics[scale=0.32]{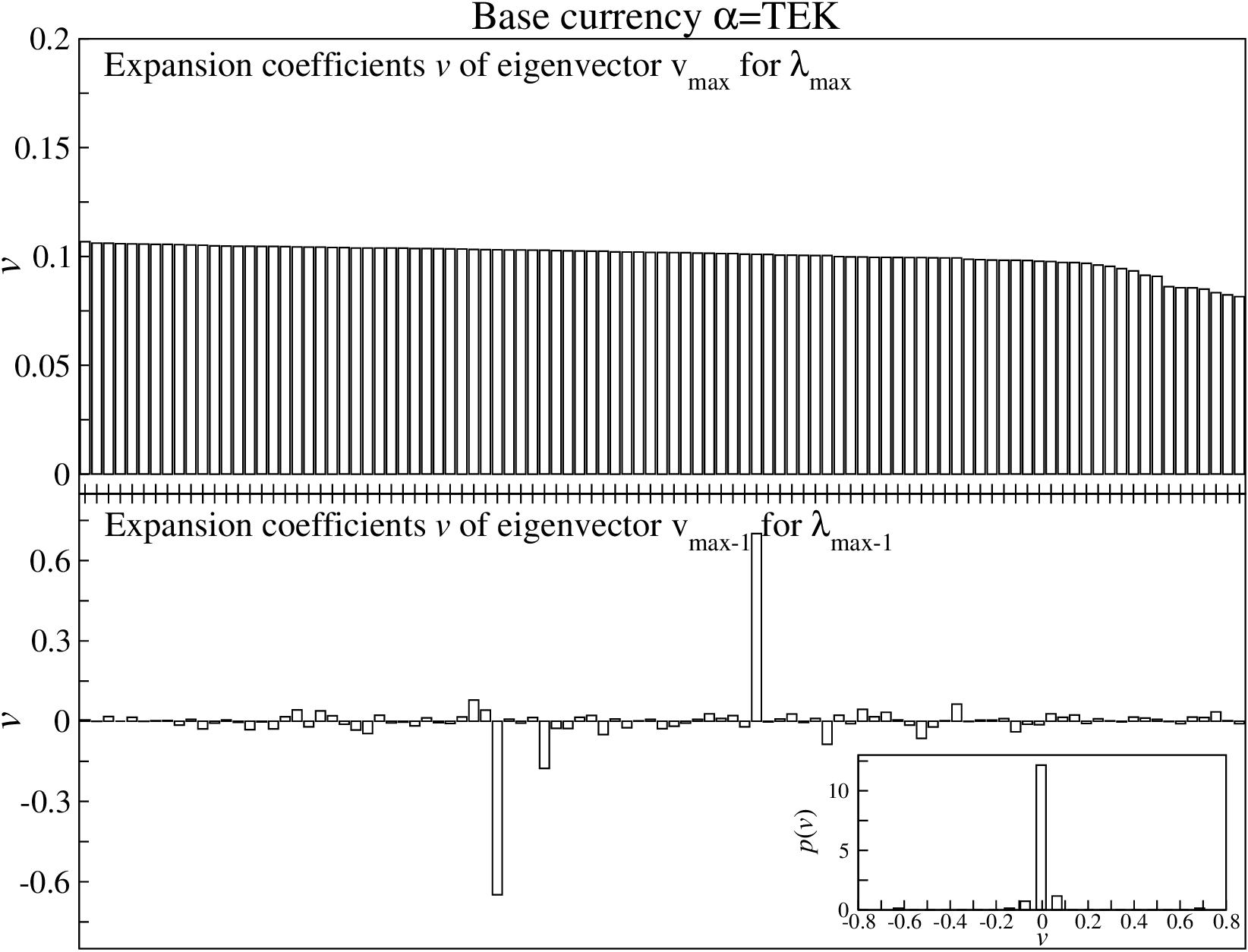}
\caption{Expansion coefficients of the eigenvectors associated with the largest and second-largest eigenvalues of the correlation matrix $C^{(\alpha)}_{ij}$, for the USD, BTC and TEK selected as a base for the remaining ones. The corresponding insets show distribution of expansion coefficients (histogram).} 
\label{fig:expansion}
\end{figure}

Complementarily, the expansion coefficients of the eigenvectors ${\bf v}_{\textrm{max}}$ corresponding to these largest eigenvalues reflect the degree of underlying collectivity. For the three cases under consideration, the expansion coefficients are shown in Fig.~\ref{fig:expansion}. By gathering a significant contribution with the same sign from all original series, they reflect the collective character of the corresponding eigenvector. Accordingly to the observed largest value of $\lambda_{\textrm{max}}$, the highest collectivity becomes apparent when TEK is taken as a base. In this case, the largest eigenvalue explains  $>90\%$ of the matrix trace, effectively `enslaving' the other eigenvalues of this correlation matrix to the region close to zero, and thus masking the noise content via compressing it~\cite{haken1987}. This effect is also seen through the expansion coefficients of the eigenvectors ${\bf v}_{\textrm{max-1}}$ associated with the second largest eigenvalue, as represented in the corresponding lower panels of Fig.~\ref{fig:expansion}. Those of the USD- and BTC-based cases already feature normal-like distributions, while the one corresponding to TEK is dominated by a restricted number of components.  

\begin{figure}
\centering
\includegraphics[scale=0.32]{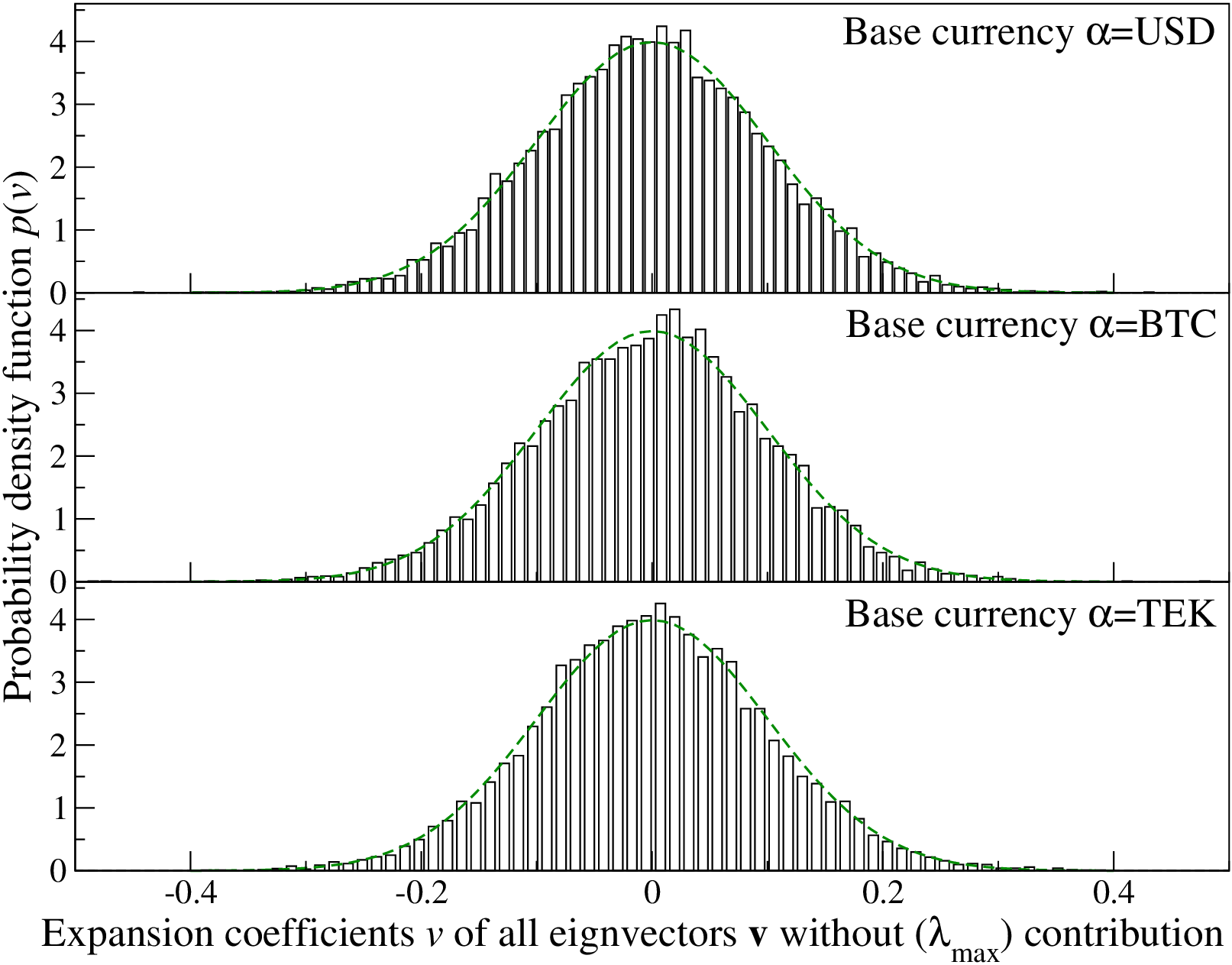} 
\caption{Distribution of expansion coefficients of all eigenvectors in the correlation matrix $C^{(\alpha)}_{ij}$, for the USD, BTC and TEK selected as a base for the remaining ones after removing the contribution of the largest eigenvalue according the prescription described in Subsection~\ref{Eigenvalue distribution} (histogram), together with the fitted Gaussian distribution (dashed line).}
\label{fig:expansion-all}
\end{figure}

To correct for such `enslaving' effects, one can remove the market factor from the data~\cite{plerou2002,kwapien2006}. This can effectively be done through least-square fitting such a factor represented by $z_{\textrm{max}}(t)$ to each of the original time-series~$g^{(\alpha)}_i(t)$:
\begin{equation}
g^{(\alpha)}_i = a_i + b_i z^{(\alpha)}_{\textrm{max}}(t) + \epsilon^{(\alpha)}_i(t), 
\label{residual}
\end{equation}
where $a_i$ and $b_i$ are parameters. One can then construct the residual correlation matrix $\bf {R}^{(\alpha)}$ from  the residuals $\epsilon^{(\alpha)}_i(t)$. Figuratively, this corresponds to viewing the positions of birds in a flock with reference to the centre-of-mass of the flock, rather than a fixed vantage point on land. The distributions of the resulting eigenvalues for the three cases of Fig.~\ref{fig:spectra} are displayed in the corresponding lower halves of the same. In all these cases, the original outlying eigenvalues disappear, and only a few remain marginally above the limits prescribed by the corresponding random Wishart matrix, while the majority fall inside these bounds. This holds even for the extreme case of considering the TEK cryptocurrency as a base. After removal of the contributions responsible for the original largest eigenvalues, the distributions of all eigenvector components in the matrix $\bf {R}^{(\alpha)}$, as shown in Fig.~\ref{fig:expansion-all}, fit well to a Gaussian, reflecting their compatibility with the random matrices.   

\subsection{Largest eigenvalues} 

\begin{figure}
\centering
\includegraphics[angle=90,origin=c,scale=0.45]{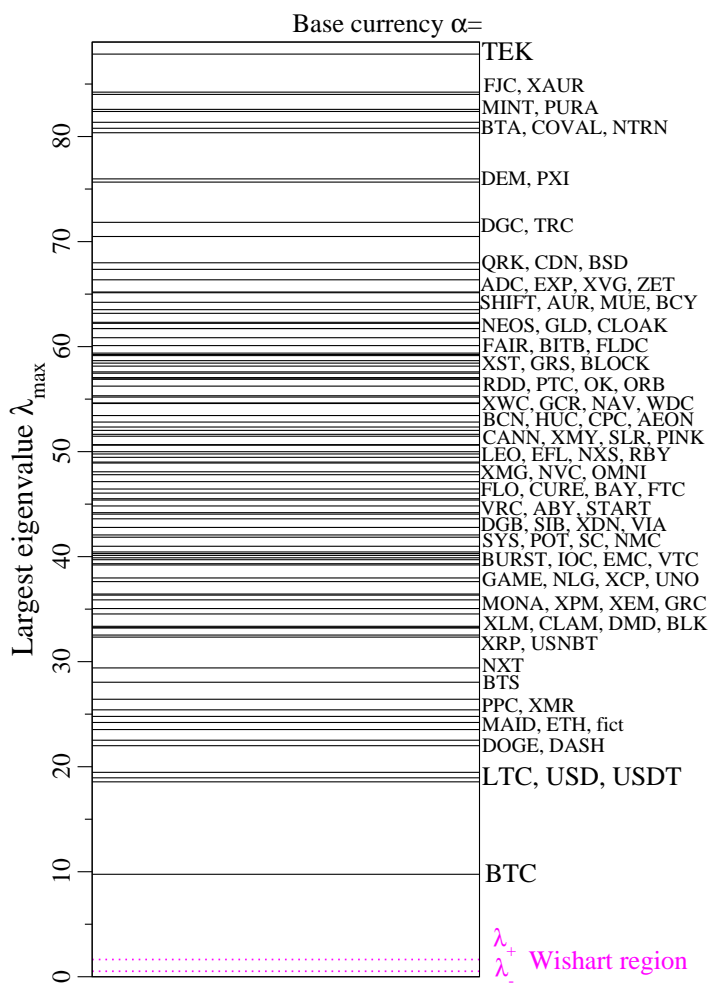} 
\caption{The largest eigenvalues in the correlation matrices $C^{(\alpha)}_{ij}$, with all 100 cryptocurrencies selectively considered as the base for the remaining ones. The USD and the fictitious cases of the base, multiplied by a factor of 99/100 to make them directly comparable, are also shown. Wishart region calculated according to Eq.~\ref{lambdaW}, assuming $\sigma_{\textrm{w}}=1$, is indicated by the dotted line.}
\label{fig:ladder}
\end{figure}

By diagonalizing the correlation matrices $\bf C^{(\alpha)}$ of all the cryptocurrencies selected as a base, one can draw the entire ladder of corresponding eigenvalues, which is displayed in Fig.~\ref{fig:ladder}; the eigenvalue of the original correlation matrix, i.e., $\alpha$=USD, is also included. The two cases of the BTC and TEK, presented in greater detail above, correspondingly constitute the lower and the upper bounds of the ladder. For the mechanisms driving the mutual exchange rates between cryptocurrencies, this is a noteworthy result. Namely, it indicates that, while involving some correlations when expressed in terms of BTC, the dynamics correlate even more markedly (yielding a larger gap) when expressed in terms of any other cryptocurrency, and, interestingly, also in terms of the USD or the fictitious currency.\par

A way to comprehend the results of Fig.~\ref{fig:ladder} is visualizing the base cryptocurrency as a reference frame for the remaining ones, and considering that this reference evolves in itself with a varying degree of independence. A change in the value of a non-influential, peripheral asset used as a reference is likely not to cause considerable changes among the values of the other currencies. From the perspective of this particular reference, their changes are thus likely to look synchronous: this results in the emergence of a well-separated eigenvalue in the corresponding correlation matrix. On the other hand, a change in the value of a significant asset, in the present case a cryptocurrency, may cause or reflect a much richer diversity of reactions among the other cryptocurrencies; consequentially, this leads to a reduction of the collectivity observed as compared to the previous case, and thus a smaller separation of the largest eigenvalue.\par

Such an interpretation finds additional legitimation when the evolution of the individual cryptocurrencies versus the remaining ones is explicitly and systematically inspected. For instance, the largest eigenvalue seen from the perspective of TEK can be attributed to its relatively small capitalization and a trend significantly different from average during the period considered, which can be seen in Fig.~\ref{fig:crypto-all}. On the other hand, it is the highest capitalization BTC, which reigned in the crypto-market by dictating the overall trend, at the same time causing diversity among smaller--amplitude fluctuations. It is also worth noting that the largest eigenvalue corresponding to the USD, the original base currency in the present basket, persists at a level noticeably higher than that of BTC. Given the above arguments, it seems natural to take this as an indication that, for the cryptocurrency market development, the BTC dynamics are already more causative even than the one of the USD. It should be noted that the largest eigenvalue corresponding to the fictitious currency, thus to the one which by construction is entirely disconnected from the crypto-market dynamics, also remains significantly above the USD, but vastly below the upper-most cases yielded by the real cryptocurrencies.

\section{Dynamics of the largest eigenvalues}

\begin{figure}
\includegraphics[scale=0.32]{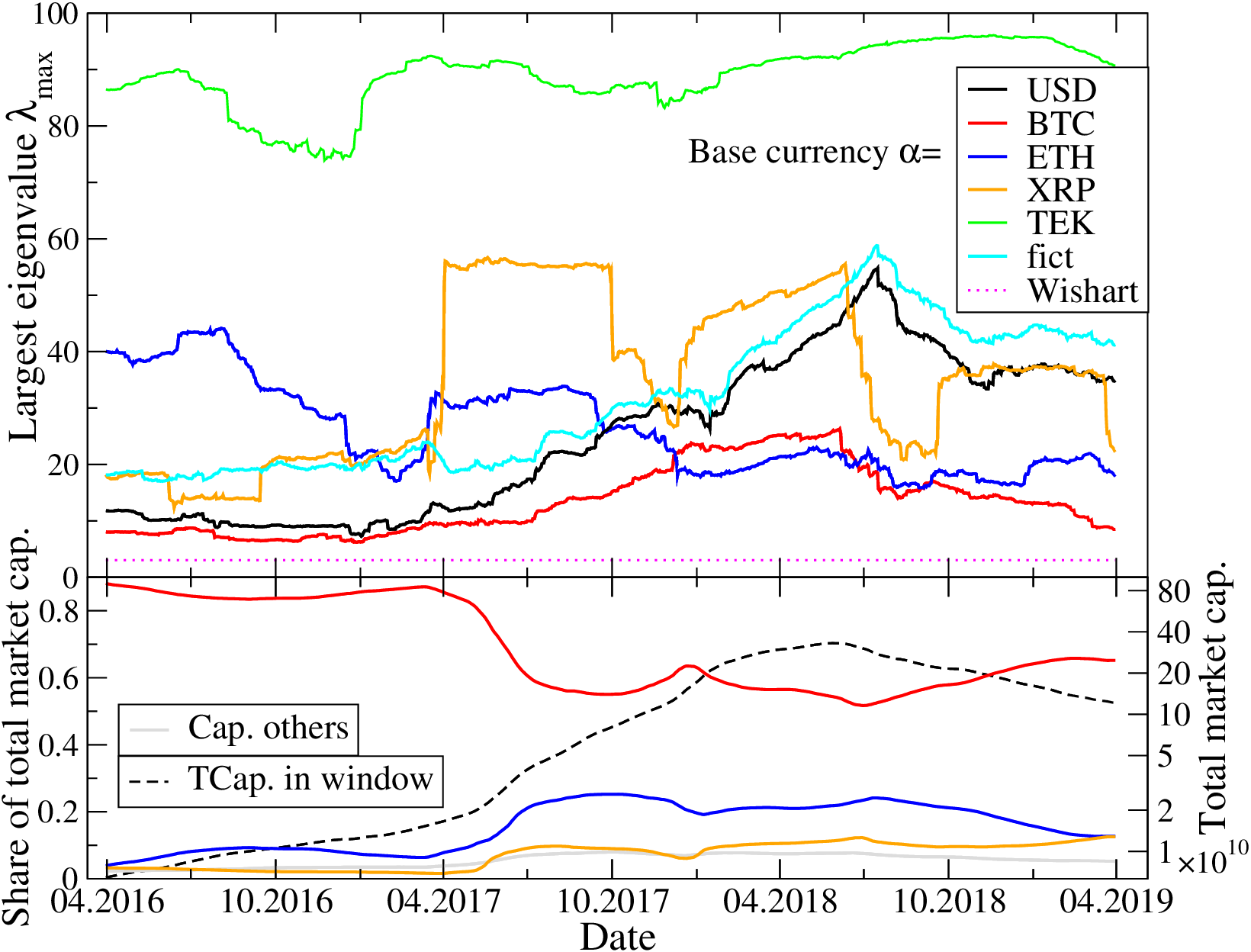} 
\caption{Upper panel: Time-development of the largest eigenvalues of the correlation matrices $C^{(\alpha)}_{ij}$, in the rolling time-window having length 182 days (half a year) moved with the step of 1 day, with BTC, ETH, XRP, TEK and USD selectively considered as the base. The fictitious case is also displayed. The corresponding Marchenko-Pastur upper limit is indicated by the dotted line. Lower panel: Total crypto-market capitalization (dashed line) within such a rolling window and corresponding share of BTC, ETH, XRP and TEK.}
\label{fig:time-dependent}
\end{figure}

The above results reveal that, within the basket of cryptocurrencies, the location of the largest eigenvalue relative to the bulk strongly depends on which cryptocurrency is used as a base, reflecting how influential that currency is. It seems natural that it is the highest-capitalization BTC, which emerges as the most influential. A related question pertains to how such characteristics change in time over the study period. The corresponding time-dependences are displayed in Fig.~\ref{fig:time-dependent} for the five cases of BTC, ETH, XRP, TEK, and USD explicitly addressed before. The correlation matrix is here calculated in a rolling time-window of length 182 days (half a year) moved with a step of 1 day. Such a length of time-window is sufficiently large $(Q=182/99)$ to avoid zero-mode degeneracies~\cite{janik2003}; the dates reported in Fig.~\ref{fig:time-dependent} corresponds to the latest day included.\par

One thus sees that the related changes in the largest eigenvalue can be sudden. The BTC and TEK, the two extremes in Fig.~\ref{fig:ladder}, largely preserve their relative locations over this rolling time-window. The BTC, however, has experienced some competition from the ETH, whose largest eigenvalue in the first half of the year 2018 decreased even below that of the BTC. Interestingly, this covers the period when the BTC price was sharply declining, and the ETH continued increasing in both price and share of the total crypto-market capitalization. This provides a further indication that the stronger a currency used as a base is, the lower the corresponding largest eigenvalue of the resulting correlation matrix is. It is also notable that towards the beginning of the study period, the largest eigenvalues when either BTC or USD are used as a base, remain close to each other, but later on, the one in the USD visibly separates by moving upwards.\par

Furthermore, in the USD-based case, the development of this largest eigenvalue starts performing moves very similar to the fictitious currency case. This can be interpreted as an indication that the cryptocurrency market dynamics become more and more disconnected from the Foreign exchange market. As a further example, in Fig.~\ref{fig:time-dependent}, the case of XRP shows somewhat peculiar sudden changes of the largest eigenvalue when this cryptocurrency is used as a base. These changes appear correlated with exceptionally large positive jumps of the XRP price relative to other cryptocurrencies, like its $102\%$ appreciation on April 2, 2017. 

\section{Concluding remarks}

As previously reported~\cite{drozdz2018}, in the year 2018, the Bitcoin (BTC) market over the years 2016-2017 attained the statistical hallmarks which are characteristic of all ``mature'' markets such as stocks, commodities and the Forex. An even more recent study \cite{drozdz2019} confirmed this observation for the year 2018 and extended it to Ethereum (ETH), another cryptocurrency which, in terms of the capitalization involved, is the second most important one among the hundreds of traded cryptocurrencies. That study revealed that the cross-correlations between the BTC/ETH and the EUR/USD exchange rates start vanishing, signaling the emergence of a disconnected crypto-market. The present work, predicated on the correlation matrix formalism, explicitly addressed the large set of 100 most liquid cryptocurrencies. The results show that this basket parallels the mature Forex market both in its random matrix-related characteristics and, most importantly, in a subtle competition between heterogeneity and collectivity among exchange rates, which elevates one large eigenvalue above the random bulk in the same way as observed for the Forex~\cite{drozdz2007,kwapien2012}. The latter effect allows identifying the most influential cryptocurrency, and it is the Bitcoin that on the crypto-market comes out as such and thus plays a similar role as the USD in the Forex~\cite{drozdz2007}. Meanwhile, on the crypto-market, the USD dynamic starts resembling a fictitious currency. All these facts point to the anticipated~\cite{drozdz2018} onset of complete disconnection of the cryptocurrency from the conventional markets. This means that not only the Bitcoin but eventually the whole crypto-market already will offer an alternative to traditional markets, effectively representing a new territory for possible portfolio diversification~\cite{guesmi2019}.

\end{document}